# Detection and Prevention Against Poisoning Attacks in Federated Learning

Viktor Valadi
AI Sweden
Gothenburg, Sweden
vviktor@student.chalmers.se

Madeleine Englund
AI Sweden
Gothenburg, Sweden
maen0191@student.umu.se

Mark Spanier
Beacom College
Dakota State University
Madison, SD, USA
mark.spanier@dsu.edu

Austin O'Brien
Beacom College
Dakota State University
Madison, SD, USA
austin.obrien@dsu.edu

## ABSTRACT

With a rapid increase in new technology, follows new cybersecurity challenges. Federated Learning is a new type of Machine Learning, first proposed in 2016. End clients work together with or without a central server to train a Machine Learning model. A Federated Learning model trains sub-models on each client and later sends the model parameters back and forth, either between each other or the central server. This enables companies and organizations to work together without sending any of their private data. Much research has gone into detecting cyberthreats in this new type of Machine Learning; however, there is a lack of research on how to prevent cyberthreats from affecting the global model. This paper proposes and investigates a new approach for detecting and preventing several different types of poisoning attacks from affecting a centralized Federated Learning model via average accuracy deviation detection (AADD). By comparing each client's accuracy to all clients' average accuracy, AADD detect clients with an accuracy deviation. The implementation is further able to blacklist clients that are considered poisoned, securing the global model from being affected by the poisoned nodes. The proposed implementation shows promising results at detecting poisoned clients and preventing the global model's accuracy from deteriorating.

## 1 Introduction

Federated Learning (FL) is a subtype of Machine Learning (ML) presented by Google in 2016 [1], where the main difference between FL and traditional ML being FL enables decentralized learning contrary to traditional ML which uses centralized learning [1-3]. Contrary to traditional ML, FL makes use of several different edge nodes, or clients, to train local sub-models [1, 4]. In a centralized FL topology, the parameters of the local sub-models in the clients are sent to a central server where they are aggregated [4]. The server thereafter sends the parameters of the global model back to the clients where the sub-models are updated [3]. The process is illustrated in Fig. 1. A decentralized FL topology does not make use of any central server [5]. Instead, the clients send their parameters directly to the nearby clients [5], as illustrated in Fig 2. Since the local models only pass on their model parameters, no data transferring is used in FL, making it a preferred alternative when working with sensitive data and/or low bandwidth [3, 4].

### 1.1 Applications

Federated Learning can be used in many different applications. However, it can be a game-changer for applications where data sharing is restricted or simply not possible, often due to legal restrictions or low bandwidth [4].

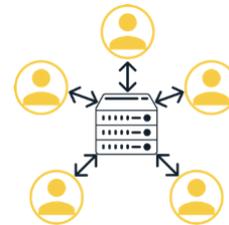

**Figure 1: Centralized Federated Learning topology**

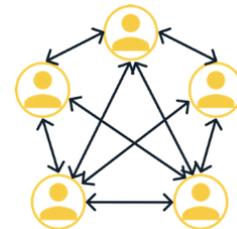

**Figure 2: Decentralized Federated Learning topology**

A few examples of such applications are autonomous vehicles [4], healthcare [6], and Low Earth Orbit (LEO) satellites [7].

*1.1.1 Autonomous Vehicles:* Research on autonomous vehicles is a rapidly developing field, where many resources are focused on how the vehicle is to behave in traffic, i.e., lane following, adaptive cruise control, navigation, etc. [8]. Currently, research on autonomous vehicles is mostly performed using traditional centralized ML where the global model is trained using sensor data collected by the car and later pushed back to the vehicle to perform its decisions [4]. Using this method to train the global model can be expensive regarding bandwidth, along with a high risk of violating the user's private data [4, 9]. An autonomous vehicle used to collect data can collect as much data as 40TB per hour [10]. In addition, it is not unusual for companies who handle autonomous vehicle data to simply transport the computer used in the car instead of sending the data to another server due to the limited bandwidth [11].

However, utilizing FL can eliminate these costs and risks [4, 8]. By only sending the local model parameters instead of the massive amount of data used to train autonomous vehicle models [4], sensitive data is no longer shared and limited bandwidth is no longer pushed to limits [4, 8].

*1.1.2 Healthcare:* ML is a highly valuable method to use within healthcare and has been used in numerous applications [12]; however, it might be most famous for its ability to detect and classify cancer in an early stage [13]. For an ML model to make the correct healthcare predictions it needs access to countless personal medical records [6]. Although, in many cases, the AI model needs more data than a single medical provider organization can produce [13]. In addition, legal restrictions hinder medical provider organizations from sharing data with each other [13, 14]. Using FL in healthcare, such as for cancer diagnostics, would allow medical provider organizations to share their local model parameters with each other, consequently, it is possible to train an FL model that can make accurate cancer predictions without sharing their patients' sensitive data [6].

*1.1.3 Low Earth Orbit Satellites:* One of the Low Earth Orbit's (LEO) main purposes is to collect earth observations to detect global occurrences, such as natural disasters or climate change [7, 15]. Most LEOs send their data to machine learning models on Earth through ground stations [15]. However, LEO satellites only come in contact with such ground stations a few times a day, resulting in the models on the ground not having access to real-time data from the satellites. Besides, this data transferring process is slow due to limited bandwidth [15]. It has been proposed to use onboard FL on LEO satellites to eliminate the problem of low bandwidth [15]. This way, only the most relevant data is sent down to the ground stations, which facilitates the transmission process when working with limited bandwidth [15].

## 1.2 Cybersecurity in Federated Learning

Cybersecurity is highly important when working with sensitive data, and is connected to vehicles, healthcare, or LEO satellites, and even though the FL model does not share the data used by the clients, the FL model is exposed to several different cyberthreats, e.g., data poisoning, and model poisoning [16, 17].

The most common poisoning attacks on an FL model are either label poisoning, which is part of data poisoning, or model poisoning [17]. In both cases, the attacker may attempt to deteriorate the global model [17]. In terms of label poisoning, the most used approach is label flipping [17]. The attacker swaps one class label for another, trying to get the clients to make the wrong predictions [17]. Model poisoning on the other hand poisons the clients' parameters before they are sent to the central server [18]. A targeted model poisoning attack aims for a high certainty that the global model is to miss-classify a specific set of inputs [18]. One such attack this project investigated was a model poisoning attack called lazy poisoning. A client that has been lazy poisoned in a centralized FL model does not use its own data during training [19]. Instead, the poisoned client takes the model parameters received from the central server and sends them back, exactly as they are [19]. Since the poisoned client does not use its own data in the FL model, it is possible for the owner of the client to later use their data to improve the finished model and thereby obtain a slightly superior model for themselves [19, 20].

Despite the vulnerabilities of FL, most research focuses on detection and not prevention of cyberthreats in FL models [16, 21-23]. This paper, therefore, aims at investigating if it is possible to detect and prevent poisoned nodes from impacting a centralized Federated Learning model by implementing an average accuracy deviation detection algorithm to verify node parameters. The proposed research investigation consists of two hypotheses. Hypothesis 1; a poisoning attack is going to decrease the global model's accuracy, and hypothesis 2; it is possible to reduce the decrease in the global model's accuracy by blacklisting nodes that are considered poisoned by using the average accuracy deviation detection algorithm. The framework used in this project is Flower – A Friendly Federated Learning Framework [24, 25].

*1.2.1 Average Accuracy Deviation Detection:* Average Accuracy Deviation Detection: Deviations have been recognized as outliers, errors, or noise, in ML for a long time [26]. However, in cybersecurity, these deviations are instead called anomalies [27]. Anomaly detection is widely used for detecting cyberthreats [27-29], although, its utilization in an FL setting has been minimal. If a client in an FL model has an accuracy that is a deviation compared to the rest of the clients, it might suggest that the client has been poisoned. Consequently, this paper proposes a new strategy to detect poisoned clients in a centralized FL model, Average Accuracy Deviation Detection (AADD).

## 2 Method

## 2.1 Centralized Federated Learning Setup

In this project, a real FL setting has been simulated locally using the Flower framework [24]. In Flower, a run script generates all certificates used by the server and the clients. The run script uses a data loader to partition the data equally between clients, resulting in homogeneous data between all clients. The implementation supports a heterogenous data split among clients, however, the effects of heterogenous data are yet to be explored. The dataset used is CIFAR10, consisting of 60000 images divided into 10 different classes [30]. The dataset is divided into a training set, consisting of 50000 images, and a test and validation set, consisting of 10000 images [30]. In the run script, a parameter stating if an individual client is poisoned or not is further passed down to the clients, giving them information about their state.

The Flower framework is used to aggregate the weights of each client and send the aggregated weights back to the clients for further training. In addition, the server performs several different evaluations. The server has two evaluation methods, one that is used only by the additional analysis and the AADD that investigates the accuracy of individual labels, and one that is used by the Flower framework to evaluate each round. In our implementation, the built-in evaluation function from Flower [31] has been extended to perform more general analysis, such as individual client accuracy, variance between clients, and individual label accuracy. These analyses are performed both for the aggregated and the individual models. The server is further able to blacklist a node if the server finds the node to be poisoned. To blacklist a node, the function "reconnect_client" in Flower can be used [32]. The function disconnects a client and later reconnects the client again after a specific amount of time, specified in the function [32]. If no time for reconnection is specified, the function disconnects the client without trying to reconnect it again [32].

The clients in Flower have the main task to train the weights they receive from the server with their data and send their updated weights back to the server. Further, if any client has been flagged as poisoned in the run script, that client will poison its own data before training. Several different poisoning strategies are supported in the implementation.

## 2.2 Federated Learning Model Baseline

In our investigations, 10 clients were used. Each client has two equally sized sets of data, which are alternated between rounds. For each round, there are 10 epochs and a batch size of 128 - 10 epochs were chosen since numerous training sessions revealed that to be the best number of epochs for the model and 128 in batch size was chosen as an attempt to expedite training with the large dataset. When finding the baseline for the model, none of the 10 clients had been poisoned.

For a poisoning attack to be classified as successful the accuracy of the poisoned final model must fall between -1% to -1.5% from the baseline's accuracy. The threshold stands for how difficult it is to successfully poison the model undetected. If a higher threshold is used, e.g., -3%, it is easier to successfully poison the model without being detected, and if a lower threshold would be used, e.g., -0.5% it would be harder to successfully poison the model without being detected.

To get the final baseline, the model was run 4 times. The model accuracy averaged around 80.4%, with deviations of around 0.2%. The baseline was therefore set to 80.4%.

## 2.3 Poisoning Strategy

To fully investigate the possibility to detect poisoned nodes from impacting a centralized FL model several different poisoning strategies have been used, including data poisoning and model poisoning. For each type of poisoning, two clients were poisoned and the amount of poisoning used for each type of poisoning was based on the results from the impact analyses.

*2.3.1 Data Poisoning:* This project has investigated two types of data poisoning, label poisoning, and image poisoning.

*2.3.1.1 Label poisoning:* During random label poisoning a random index for both partitions of a client's data, i.e., an index between 0-2499 and an index between 2500-5000, is changed to a random number between 0-9. A data entry with label 1 could be changed for example to label 3. 900 out of 2500 images for each partition of the dataset were poisoned. This random label poisoning can occur for x number of indices and y number of clients, depending on how strong the poisoning should be. See appendix 1 for code example.

The specific label poisoning poisons all labels from a specific category. Two methods of poisoning have been used, specific category to specific category and specific category to random category. The specific category to specific category takes all labels from one category and changes the labels to a label belonging to another specific category. All labels are changed to the same category. The specific category to random category takes all labels from a specific category and changes the labels to a label belonging to a random category. See appendix 1 for code example.

*2.3.1.2 Image poisoning:* Image poisoning iterates over every image in each client. For each image, random pixels were chosen 600 times. For each chosen pixel, the values of each color channel are multiplied by a random number between 0.5 and 1.5. See appendix 1 for code example.

*2.3.2. Model poisoning:* In this project, the model poisoning type lazy poisoning has been investigated.

*2.3.2.1 Lazy poisoning:* If a client is poisoned by lazy poisoning, the client takes the model parameters sent from the central server and sends the same model parameters back to the central model. While the other, not poisoned, clients are updating the model parameters, the poisoned client is not contributing to improving the global model by sending the same old parameters back.

The blacklisting method was used for lazy poisoning. If a client is detected as poisoned 3 times during the last 10 rounds, the method will blacklist the client and disconnect it.

## 2.4 Impact Analysis

An impact analysis has been performed for all types of poisoning, to investigate how much poisoning is necessary for the global model's accuracy to drop below 1% of the baseline, however, not more than 1.5%. As previously stated, 10 clients have been used in the FL model, with equally sized sets of data. 16 rounds, with 10 epochs and a batch size of 128 were used. Each impact analysis was performed with two poisoned clients. Each client has a dataset of 5000 images, divided into two partitions resulting in 2500 images per round. To find the correct amount of poisoning needed, several different training sessions with different amounts of poisoning were performed. When the global model's accuracy reached the goal between -1% to -1,5% compared to the baseline, the impact analysis was considered done for the type of poisoning used. However, if the global model's accuracy did not reach the threshold goal, but instead managed to detect 1/3 of the poisoned rounds, it was considered sufficient poisoning.

## 2.5 Average Accuracy Deviation Detection

This project proposes a new method for detecting poisoned nodes in a centralized FL model, Average Accuracy Deviation Detection (AADD). Over the course of the project, two versions of AADD have been tested. These two versions will hereafter be referred to as AADD 1.0 and AADD 2.0.

AADD 1.0 looks for deviations in accuracy from the clients' individual models. This is implemented in the current version by averaging the accuracy of all clients after each round and asserting none of them are below a value given by the threshold function of the average accuracy. The current threshold function is dependent on which round it is and can be seen in Fig. 3. The threshold function was designed by looking at the variance for each training client's accuracy each training round and by trying to fit the same curve.

AADD 2.0 is an extension of AADD 1.0. In version 2.0 the same average accuracy deviation detection as in the previous version is performed. However, the detection has been extended to investigate individual labels as well. Note that this assumes homogeneity between labels, and any labels that do not exist in all clients' data should be excluded from the individual label check.

The individual label check functions the same way as AADD 1.0. AADD 2.0 calculates the accuracy of each label for every client. Each individual model's accuracy for each label is compared to the mean accuracy of the same label. If one model's accuracy is below the threshold, the client is identified as poisoned. The threshold function used for individual labels can be seen in Fig. 4. It is the same function as the previous threshold function, scaled by a factor of 4.8 as this number very rarely gave any false positives. Fig. 5 displays a table for how a single round of data may be perceived and concluded. See appendix 1 for the core code of AADD.

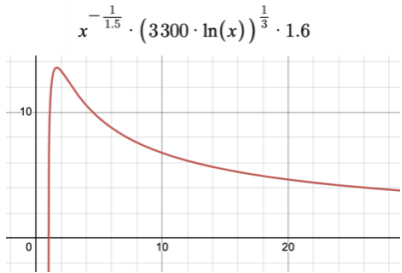

**Figure 3: AADD 1.0 threshold function (starts on x = 2).**

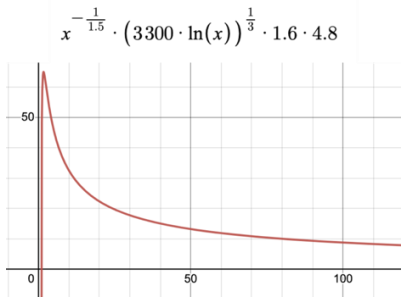

**Figure 4: AADD 2.0 threshold function (starts on x = 2).**

| Client \ Label | 0 | 1 | 2 | Avg |
|---|---|---|---|---|
| 0 | 60% | 73% | 75% | 69.3% |
| 1 | 57% | 68% | 77% | 67.3% |
| 2 | 62% | 45% | 43% | 50% |
| 3 | 22% | 67% | 84% | 57.6% |
| 4 | 59% | 70% | 65% | 64.6% |
| Avg | 52% | 64.6% | 68.6% | 61.76% |

**Figure 5: AADD 2.0 fabricated example of a training round. Red arrows indicate a poisoned client. The last row and column displaying averages, do not have the same threshold as the rest of the columns as it is part of AADD 1.0.**

## 3 Results

### 3.1 Federated Learning Model Baseline

The average accuracy for the FL model baseline is 80.4% of images predicted correctly. For each individual label, the baseline can be seen in Table 1.

### 3.2 Federated Learning Model Baseline

Results of the different impact analyses for each type of poisoning are stated below.

*3.2.1 Data Poisoning:*

*3.2.1.1 Label poisoning:* Impact analysis for random label poisoning, where 900 out of 2500 images on two clients had been poisoned, resulted in a decrease in the central model's accuracy by -1.4 percentage points. Hence, 900 out of 2500 images on two clients are considered the right amount of poisoning for our investigations.

Impact analysis for specific to specific label poisoning on two clients resulted in a decrease in the central model's accuracy by -0.06 percentage points. This model accuracy did not fall between the threshold for the poisoning to be considered enough. However, based on the confusion matrix (see table 2) the poisoning is still considered a sufficient amount of poisoning.

Impact analysis for specific to random label poisoning on two clients, resulted in a decrease in the central model's accuracy by 2.1 percentage points. Thus, it is considered enough poisoning.

*3.2.1.2 Image poisoning:* Impact analysis for image poisoning, where 2500 images of both partitions of the dataset of two clients had random pixels altered within a threshold of 50% 600 times, resulted in a decrease of 0.9 percentage points. Based on the confusion matrix, this amount of poisoning is considered enough (see table 5).

*3.2.2 Model poisoning:* Impact analysis for lazy poisoning on two clients resulted in a decrease in the central model's accuracy by 1.01 percentage points. This decrease in accuracy falls within our threshold.

### 3.3 Average Accuracy Deviation Detection

Results of the different impact analyses for each type of poisoning are stated below. Further results can be found in Appendix 2.

*3.3.1 Data Poisoning:*

*3.3.1.1 Label poisoning:* AADD 1.0 for random label poisoning resulted in 16 detected poisoning attacks, 16 missed poisoning attacks, 3 false positive poisoning attacks, and 125 non-poisoning attacks (see Table 2).

AADD 1.0 for specific to specific label poisoning resulted in 14 detected poison attacks, 18 missed poison attacks, 1 false positive poison attack, and 127 non-poison attacks (see Table 3). The specific label accuracies for the global model can be seen in Table 1.

AADD 2.0 for specific to random label poisoning resulted in 22 detected poison attacks, 10 missed poison attacks, 3 false positive poison attacks, and 125 non-poison attacks (see Table 4). The specific label accuracies for the global model can be seen in Table 1.

*3.3.1.2 Image poisoning:* AADD 1.0 for image poisoning resulted in 16 detected poison attacks, 16 missed poison attacks, 3 false positive poison attacks, and 125 non-poison attacks (see Table 5).

*3.3.2 Model Poisoning:* AADD 2.0 for lazy poisoning resulted in 6 detected poison attacks, 26 missed poison attacks, 3 false positive poison attacks, and 125 non-poison attacks (see Table 6).

**Table I.**

**Label Accuracies of Global Model**

| Label | Accuracy (%) | | | | |
|---|---|---|---|---|---|
| | Baseline | Specific to Specific Label | Specific to Random Label | Lazy Poisoning | Blacklisting Lazy Poisoning |
| 0 | 81 | 83.6 | 84.7 | 83.2 | 80 |
| 1 | 87.4 | 90.1 | 88.2 | 86.7 | 88.2 |
| 2 | 57.4 | 68.4 | 46.1 | 61.5 | 64.7 |
| 3 | 65.2 | 56.6 | 58 | 61.9 | 62.5 |
| 4 | 83.4 | 72.7 | 84.9 | 83.8 | 84 |
| 5 | 69 | 73.9 | 65 | 66.7 | 63 |
| 6 | 92.9 | 90.7 | 94.3 | 89.5 | 92.7 |
| 7 | 86 | 84.3 | 83.2 | 81.1 | 79 |
| 8 | 91.5 | 92.4 | 90 | 91.1 | 90 |
| 9 | 91.1 | 90.8 | 89 | 87.8 | 88.6 |

**Table II.**

**Confusion Matrix for Random Label Poisoning**

| N = 160 | Actual positive | Actual negative | |
|---|---|---|---|
| Classified positive | 16 | 3 | N = 19 |
| Classified negative | 16 | 125 | N = 141 |
| | N = 32 | N = 128 | |

**Table III.**

**Confusion Matrix for Specific to Specific Label Poisoning**

| N = 160 | Actual positive | Actual negative | |
|---|---|---|---|
| Classified positive | 14 | 1 | N = 15 |
| Classified negative | 18 | 127 | N = 145 |
| | N = 32 | N = 128 | |

**Table IV.**

**Confusion Matrix for Specific to Random Label Poisoning**

| N = 160 | Actual positive | Actual negative | |
|---|---|---|---|
| Classified positive | 22 | 3 | N = 25 |
| Classified negative | 10 | 125 | N = 135 |
| | N = 32 | N = 128 | |

**Table V.**

**Confusion Matrix for Image Poisoning**

| N = 160 | Actual positive | Actual negative | |
|---|---|---|---|
| Classified positive | 16 | 3 | N = 19 |
| Classified negative | 16 | 125 | N = 141 |
| | N = 32 | N = 128 | |

**Table VI.**

**Confusion Matrix for Lazy Poisoning**

| N = 160 | Actual positive | Actual negative | |
|---|---|---|---|
| Classified positive | 6 | 3 | N = 9 |
| Classified negative | 26 | 125 | N = 151 |
| | N = 32 | N = 128 | |

**Table VII.**

**Confusion Matrix for Lazy Poisoning with Blacklisting Implemented**

| N = 134 | Actual positive | Actual negative | |
|---|---|---|---|
| Classified positive | 6 | 3 | N = 9 |
| Classified negative | 0 | 125 | N = 151 |
| | N = 6 | N = 128 | |

The specific label accuracies for the global model can be seen in Table 1.

Blacklisting poisoned nodes resulted in the global model's accuracy decreasing by 1.14 percentage points. The AADD when using the blacklisting function resulted in 6 detected poison attacks, 0 missed attacks, 3 false positives, and 125 non-poison attacks (see Table 7).

## 4 Discussion

### 4.1 Method

Most of the poisoning methods have the risk of overlaps, for example, if label one was changed to label one, or the same data was changed more than once. This means that when we claim 600 labels were changed at random, the actual number of labels changed would most likely be somewhere around 530 labels changed.

### 4.2 Average Accuracy Deviation Detection

The proposed method for detecting poisoned clients is still a work in progress. The algorithm has only been tested in one setting currently, so more testing is required to ensure its robustness.

There are most likely two weak links currently. First the threshold function and second working on heterogeneous data. We would propose a function dependent on how much the global model accuracy increased in the last round, the variance of the clients' accuracy, and the individual clients' accuracy, instead of the current function that only depends on the current round. If no such function is available a less steep threshold function would most likely work on most edge learning settings. The other weak link is as mentioned working on heterogeneous data. No investigation of detection on heterogeneous data has been performed and the algorithm would likely need an extension to work with clients poisoning data only available to a smaller subset of all clients.

Even with these weak links, we believe the current algorithm is a great tool for protection against poisoning attacks. At the beginning of the project, a static threshold function was used. This static threshold function was still able to protect against any severe poisoning attack. A static threshold function would most likely work in any federated learning setting.

### 4.3 Results

The results look promising for AADD. More sophisticated poisoning methods should be investigated. With sufficient test data for the algorithm, AADD should detect most poisoning attacks or corrupted nodes. Lastly, lazy poisoning was not able to detect the poisoning according to our own benchmarks. A proposed solution to handle lazy poisoning could be placed in the training step as opposed to in the detection part. If more training is done by the clients each round, then the gap between the client(s) performing lazy poisoning accuracy and other clients would be greater, enabling the lazy poisoning clients to be detected by the detection methods. For example, this could be done by simply increasing the number of epochs each client trains on.

### 4.4 Average Accuracy Deviation Detection in a Real Federated Learning Setting

In a real FL setting, as for training on cars, for example, we suggest that the analysis with AADD is run in parallel with the training done on the edge nodes. The total run time for the analysis using AADD depends on the size of the test data and the number of clients. Each client's individual model will be applied and measured on the test dataset. In the current setting, with 10000 test images and 10 clients, the analysis takes around 3-4 minutes to perform. With more clients and a larger test set, the analysis will take more time. To avoid this, we suggest starting by aggregating the weights and sending them back to the edge nodes, while the edge nodes continue the training the server can work on analyzing the individual models. The next weights from a compromised node can then be thrown out on arrival.

### 4.5 Blacklisting clients

As showcased in the results one attempt to blacklist two clients that were lazy poisoned was performed. The clients were blacklisted in the third round of training and did not contribute to the global model after this. This still resulted in a loss in accuracy for the final model, most likely because more training would be needed to compensate for the missing data from the clients.

## 5 Conclusion

Based on the test where two poisoned nodes were blacklisted after being detected three times, the conclusion was made that blacklisting these clients would not improve the global model. This could be explained by the lost training data when blacklisting a client. Because of this, we claim that AADD in this setting is sufficient to protect the global model against the poisoning attempts investigated.

Another conclusion is that, like in most cybersecurity fields, if you are creative enough, there is always a possibility to bypass security measures. This project has been an iterative process of finding new creative ways to bypass the security measures implemented and extending the AADD algorithm to detect the new poisoning strategy. We believe it is a good idea to start with a good foundation of security against poisoning attacks in federated learning, making it difficult to bypass any intrusion detection system.

## 6 Future Research

So far, these versions of AADD, i.e., AADD 1.0 and AADD 2.0, look for deviations in a client's average accuracy and individual label accuracy. We believe an extension that further looks at accumulated deviation from the average for both individual labels and average accuracy on the clients' individual models over all rounds would reduce the risk of a decrease in accuracy of the global model. This method would use another threshold function similar to the other detections used so far.

A few suggestions for how the threshold slopes could be improved to be more robust have been discussed but not yet explored. The main goal is to remove the dependency on rounds and move it to be dependent only on the data received from the model. Two ideas for this implementation are suggested. The easier one is to only use the accuracy of a large partition of the clients whose accuracies are closest to the average accuracy and calculate the mean and variance of those to see if any client is out by a factor of the standard deviation of the mean value. This way outliers are excluded and therefore, possible poisoned nodes.

The second, more robust approach, is to train a simple ML model, e.g., logistic regression, with a few parameters from the global model. These parameters could, for example, be mean, variance, and how much the global model accuracy increased from the previous round. To train this ML model, data gathered from simulated poisoning attacks on different types of models would be used.

In addition to these approaches, there are large overarching areas that one needs to consider. First, the effect of multiple clients working together to deteriorate the global model, second, when heterogeneous data is used, both in a centralized and decentralized FL setting, third, further research is needed in different settings of FL. Larger and more difficult problems must be investigated to understand how they differ from smaller settings.

## ACKNOWLEDGMENT

This project has been supported by AI Sweden and Dakota State University. We thank our supervisors Dr. Mark Spanier, Dr. Austin O'Brien from Dakota State University, Dr. Mats Hanson, and Dr. Mats Nordlund from AI Sweden for their guidance throughout this project. We further thank all partners of AI Sweden who have participated in discussions throughout the project. We thank Dr. Johan Östman, Dr. Mina Alibeigi, and Kim Henriksson from AI Sweden for all the support. And lastly, we thank Charles Novak for always supporting us.

# APPENDIX 1

## 1. Code Examples

```python
def poison_random_label(self,y_train,no_labels=800):
    print("Poisoning labels!!!!!!!!")
    print(len(y_train))
    y_train = list(y_train)
    for i in range(no_labels):
        x = [0.0 for j in range(10)]
        x[random.randint(0,9)] = 1.0
        y_train[random.randint(0,int(len(y_train)/2))-1] = np.array(x)
        y_train[random.randint(int(len(y_train)/2),len(y_train)-1)] = np.array(x)
    y_train = np.array(y_train)
    return y_train
```

Figure 6: Random label poisoning code

```python
def poison_specific_label(self,y_train,part_of_labels=1.0, label=4, to_label=5):
    print("Poisoning labels!!!!!!!!")
    print(len(y_train))
    y_train = list(y_train)
    for i in range(len(y_train)):
        if np.argmax(y_train[i]) == label:
            if random.uniform(0,1) <= part_of_labels:
                x = [0.0 for j in range(10)]
                if to_label == 'random':
                    to_label = random.randint(0,9)
                x[to_label] = 1.0
                y_train[i] = x
    y_train = np.array(y_train)
    return y_train
```

Figure 7: Specific label poisoning code

```python
def poison_random_pixels(self, x_train, perc_img=1.0, nr_pixels = 600, th = 0.5):
    print("Poisoning pixels!!!!!!!!")
    x_train = list(x_train)
    nr_pictures = int(perc_img*len(x_train))
    index_value = random.sample(list(enumerate(x_train)), nr_pictures)
    for idx, _ in index_value:
        for i in range(nr_pixels):
            position = random.randint(0,len(x_train[idx])-1)
            row_position = random.randint(0,len(x_train[idx][position])-1)
            x_train[idx][position][row_position]
            for j in range(len(x_train[idx][position][row_position])):
                current = x_train[idx][position][row_position][j]
                new = np.float32(round(random.uniform(current*(1-th),current*(1+th))))
                x_train[idx][position][row_position][j] = new
    x_train = np.array(x_train)
    return x_train
```

Figure 8: Random pixel poisoning code

```python
def check_poison(self,acc_new,acc_avg, scale = 1):
    th = self.epsilon(scale)
    return((acc_avg-th) > acc_new)

def epsilon(self,scale):
    x = self.round + 2
    return ((x**(-1/1.5))*((3300*np.log(x))**(1/3))*1.6*scale)/100
```

Figure 9: AADD core code

# APPENDIX 2

## 1. Further Poisoning Results

**Table VIII.**

**Result for Image Poisoning**

| Tot clients | Poisoned clients | Images/client | Rounds | Epochs | Poisoning strategy | Poisoned images | Accuracy not poisoned (%) | Accuracy poisoned (%) |
|---|---|---|---|---|---|---|---|---|
| 10 | 2 | 5000 in tot<br><br>2500 images per round | 16 | 10 | image poisoning | 5000 + 5000 pictures poisoned, of the 50k dataset.<br><br>600 pixels changed randomly, within a threshold of 50% | 80.4 | 79.5 |

**Table IX.**

**Result for Random Label Poisoning**

| Tot clients | Poisoned clients | Images/client | Rounds | Epochs | Poisoning strategy | Poisoned images | Accuracy not poisoned (%) | Accuracy poisoned (%) |
|---|---|---|---|---|---|---|---|---|
| 10 | 2 | 5000 in tot<br><br>2500 images per round | 16 | 10 | Random label poisoning | 900/2500 | 80.4 | 79.5 |

**Table X.**

**Confusion Matrix for Image Poisoning**

| N = 160 | Actual positive | Actual negative | |
|---|---|---|---|
| **Classified positive** | 24 | 1 | N = 25 |
| **Classified negative** | 8 | 127 | N = 135 |
| | N = 32 | N = 128 | |

Additional notes: Except for the first round, 4 consecutive rounds of poisoning in the earliest training phase were missed, i.e., rounds 2-5. AADD 1.0 was used.

**Table XI.**

**Confusion Matrix for Image Poisoning**

| N = 160 | Actual positive | Actual negative | |
|---|---|---|---|
| **Classified positive** | 16 | 3 | N = 19 |
| **Classified negative** | 16 | 125 | N = 141 |
| | N = 32 | N = 128 | |

Additional notes: AADD 1.0 was used.

### Table XII.
### Result for Random Label Poisoning

| Tot clients | Poisoned clients | Images/client | Rounds | Epochs | Poisoning strategy | Poisoned images | Accuracy not poisoned (%) | Accuracy poisoned (%) |
|---|---|---|---|---|---|---|---|---|
| 10 | 2 | 5000 in tot<br>2500 images per round | 16 | 10 | Random label poisoning | 800/2500 | 80.4 | 79.7 |

### Table XIII.
### Result for Specific to Specific Label Poisoning

| Tot clients | Poisoned clients | Images/client | Rounds | Epochs | Poisoning strategy | Poisoned images | Accuracy not poisoned (%) | Accuracy poisoned (%) |
|---|---|---|---|---|---|---|---|---|
| 10 | 2 | 5000 in tot<br>2500 images per round | 16 | 10 | Specific to random label poisoning | All images of 1 label moved to a random label | 80.4 | 79.3 |

### Table XIV.
### Confusion Matrix for Random Label Poisoning

| N = 160 | Actual positive | Actual negative | |
|---|---|---|---|
| **Classified positive** | 10 | 0 | N = 10 |
| **Classified negative** | 22 | 128 | N = 150 |
| | N = 22 | N = 128 | |

Additional notes: Accuracy is not within the threshold. AADD 1.0 was used.

### Table XV.
### Confusion Matrix for Specific to Specific Label Poisoning

| N = 160 | Actual positive | Actual negative | |
|---|---|---|---|
| **Classified positive** | 14 | 1 | N = 15 |
| **Classified negative** | 18 | 127 | N = 145 |
| | N = 32 | N = 128 | |

**Table XVI.**

**Result for Specific to Random Label Poisoning using AADD 1.0**

| Tot clients | Poisoned clients | Images/client | Rounds | Epochs | Poisoning strategy | Poisoned images | Accuracy not poisoned (%) | Accuracy poisoned (%) |
|---|---|---|---|---|---|---|---|---|
| 10 | 2 | 5000 in tot<br><br>2500 images per round | 16 | 10 | Specific to random label poisoning | All images of 1 label moved to a random label | 80.4 | 79.3 |

**Table XVII.**

**Result for Specific to Random Label Poisoning using AADD 2.0**

| Tot clients | Poisoned clients | Images/client | Rounds | Epochs | Poisoning strategy | Poisoned images | Accuracy not poisoned (%) | Accuracy poisoned (%) |
|---|---|---|---|---|---|---|---|---|
| 10 | 2 | 5000 in tot<br><br>2500 images per round | 16 | 10 | Specific to random label poisoning | All images of 1 label moved to a random label | 80.4 | 78.3 |

**Table XVIII.**

**Confusion Matrix for Specific to Random Label Poisoning using AADD 1.0**

| N = 160 | Actual positive | Actual negative | |
|---|---|---|---|
| **Classified positive** | 6 | 3 | N = 9 |
| **Classified negative** | 26 | 125 | N = 151 |
| | N = 32 | N = 128 | |

Additional notes: Label 2 was poisoned to a random label. AADD 1.0 was used.

**Table XIX.**

**Confusion Matrix for Specific to Random Label Poisoning using AADD 2.0**

| N = 160 | Actual positive | Actual negative | |
|---|---|---|---|
| **Classified positive** | 22 | 3 | N = 25 |
| **Classified negative** | 10 | 125 | N = 135 |
| | N = 32 | N = 128 | |

Additional notes: Scale of threshold = 4. Label 2 poisoned. AADD 2.0 was used.

**Table XX.**

**Result for Lazy Poisoning**

| Tot clients | Poisoned clients | Images/client | Rounds | Epochs | Poisoning strategy | Poisoned images | Accuracy not poisoned (%) | Accuracy poisoned (%) |
|---|---|---|---|---|---|---|---|---|
| 10 | 2 | 5000 in tot<br><br>2500 images per round | 16 | 10 | Lazy poisoning | No images were used for training | 80.4 | 79.3 |

**Table XX.**

**Confusion Matrix for Lazy Poisoning**

| N = 160 | Actual positive | Actual negative | |
|---|---|---|---|
| **Classified positive** | 6 | 3 | N = 9 |
| **Classified negative** | 26 | 125 | N = 151 |
| | N = 32 | N = 128 | |

Additional notes: scale of threshold = 4.8. Lazy poisoning managed to decrease the accuracy with more than 1% while still not being detected well. Only in the first 3 rounds were we able to detect it.